\pdfoutput=1

\documentclass[11pt]{article}

\usepackage[final]{acl}

\usepackage{times}
\usepackage{latexsym}

\usepackage[T1]{fontenc}

\usepackage[utf8]{inputenc}
 \usepackage{tabularray}

\usepackage{microtype}

\usepackage{inconsolata}

\usepackage{graphicx}
\graphicspath{{./images/}}

\usepackage{listings}
\usepackage{color}
\usepackage{subcaption}
\definecolor{codegreen}{rgb}{0,0.6,0}
\definecolor{codegray}{rgb}{0.5,0.5,0.5}
\definecolor{codepurple}{rgb}{0.58,0,0.82}
\usepackage{todonotes}
\title{Multi Agent based Medical Assistant for Edge Devices}

\author{Sakharam Gawade\thanks{Equal Contribution}, Shivam Akhouri\footnotemark[1],  Chinmay Kulkarni\footnotemark[1], Jagdish Samant\footnotemark[1], Pragya Sahu\footnotemark[1], \\ {\bf Aastik, Jai Pahal, Saswat Meher }\\
	Samsung Research Institute Bangalore, India\\
	Team Name: Agent Lite
}

\begin{document}
	\maketitle
	\begin{abstract}
		Large Action Models (LAMs) have revolutionized intelligent automation, but their application in healthcare faces challenges due to privacy concerns, latency, and dependency on internet access. This report introduces an on-device, multi-agent healthcare assistant that overcomes these limitations. The system utilizes smaller, task-specific agents to optimize resources, ensure scalability and high performance. Our proposed system acts as a one-stop solution for health care needs with features like appointment booking, health monitoring, medication reminders, and daily health reporting. Powered by the Qwen Code Instruct 2.5 7B model, the Planner and Caller Agents achieve an average RougeL score of 85.5 for planning and 96.5 for calling for our tasks while being lightweight for on-device deployment. This innovative approach combines the benefits of on-device systems with multi-agent architectures, paving the way for user-centric healthcare solutions. Check our code \href{https://github.com/sakharamg/Multi-Agent-Health-Assistant/}{here}.
	\end{abstract}
	
	\section{Introduction}
	Large Action Models (LAMs) have emerged as a groundbreaking paradigm in intelligent automation, enabling systems to execute complex tasks by dividing them into modular, agent-specific components. These systems excel in domains like e-commerce and logistics, where cloud-based architectures leverage vast computational resources to deliver high precision and scalability. However, applying this framework to healthcare poses unique challenges that cloud-centric systems are ill-suited to address.
	
	Healthcare applications require handling sensitive personal data, ensuring real-time responsiveness, and maintaining uninterrupted functionality, even in offline scenarios. Cloud-based solutions, while powerful, are often constrained by privacy risks, latency issues, and dependency on continuous internet access. These challenges necessitate a shift toward on-device implementations, where user data remains \textbf{private}, and operations are optimized for local execution.
	
	The computational limitations of edge devices pose an additional challenge when deploying large, monolithic models. This is where the power of multi-agent systems comes into play. By utilizing smaller, task-specific agents, we can effectively \textbf{distribute workloads}. This leads to optimized\textbf{ resource allocation}, where compact models are designed to operate within the constraints of edge devices, ensuring efficient use of available computational resources. Moreover, the system's functionality can be easily \textbf{scaled} by introducing additional agents, allowing for incremental growth and adaptation. This setup fosters ``modular collaboration'', where each agent functions independently, focusing on its specific task, while also working in harmony with others to manage more complex workflows. 
	
	Our solution uses a multi-agent architecture with on-device data and specialized models for intelligent diagnosis, appointment booking, emergency services, etc. It is integrated with a user-friendly application, enabling seamless interaction and personalized data retrieval, enhancing agent capabilities through retrieval augmented generation.

	
	
	
In this project, we contribute the following:
\begin{enumerate}
	\item A multi-agent healthcare assistant designed for \textbf{edge devices}, ensuring \textbf{privacy} and \textbf{scalability} (refer section \ref{sec:sys-arch}).
	\item A \textbf{data creation pipeline} that generates \textbf{synthetic data} and enhances it to address challenges relevant to AI agents (refer section \ref{sec:datagen}).
	\item The fine-tuned planner and caller agents that achieve an average RougeL score of \textbf{85.5 for planning and 96.5 for calling} using our generated data, demonstrating effectiveness for our target use cases (refer to section \ref{sec:results}).
	\item An \textbf{application} that connects users with the multi-agent system, enabling \textbf{personalized data interaction} (refer section \ref{sec:app-design}).
\end{enumerate}

	\section{Motivation}
	In the midst of today's hectic lifestyle, staying on top of one's health can feel like an uphill battle. Juggling medications, monitoring vital signs, and keeping track of appointments can easily become overwhelming, especially for vulnerable groups such as children and the elderly, who often require more frequent medical attention. Our multi-agent healthcare assistant is designed to simplify and streamline this process, empowering users of all ages to take charge of their health journey while ensuring the utmost privacy. By harnessing the power of local, on-device processing and seamless integration with smartwatches, we offer a secure, user-friendly solution that makes health management efficient and accessible for everyone, regardless of age or technological expertise.
	
	\section{Problem Statement and Use-cases}
	\label{ps_and_usecases}
	We develop a multi-agent system that manages personal health data securely and efficiently while providing users with timely reminders, monitoring vitals, and handling appointment requests based on symptoms. It operates locally on the user's device, ensuring data privacy and seamless integration with smart watches for enhanced user experience.
	
	
	The primary usecases for the project are:
	
	\textbf{Intelligent Diagnosis and Appointment Scheduling:}
	The healthcare assistant engages in multi-turn conversations with the user, allowing for intelligent diagnosis by gathering comprehensive information about the user's health concerns. This iterative questioning process is key to accurately identifying the most appropriate medical specialist, ensuring targeted and effective care. It also uses past interactions and profile information to identify the right specialist. The aim is not to diagnose or provide remedies, but to identify the most suitable specialist based on the user's needs and availability. This end-to-end functionality reduces user burden and ensures timely consultations without unnecessary hassle.
	
	\textbf{Emergency SOS:}
	The system is equipped to handle critical emergency situations where immediate assistance is required. Users can manually trigger an SOS signal, which prompts the system to take swift action. We call this as \underline{Hard SOS} where we are sure of an emergency situation. These actions may include notifying emergency services (e.g., calling an ambulance), contacting the user’s designated emergency contacts, and providing real-time updates to responders. The system is designed to ensure prompt escalation of emergencies, enhancing the user's safety and response time during life-threatening scenarios.
	
	\textbf{Vitals Tracking and Abnormality Detection:}
	The system continuously monitors the user’s vital signs through smart devices like wearable health trackers or smartphones. This feature ensures proactive health management by identifying anomalies or irregularities in real time, such as abnormal heart rate, irregular blood oxygen levels, or deviations in other key health metrics. When a abnormal situation is detected, a SOS is triggered, alerting the user and providing actionable recommendations. We call it \underline{Soft SOS} as unlike Hard SOS, it only alerts the user. For this project, we integrate with a simulator smart watch.

	\textbf{Schedule Reminders:}
	Managing healthcare routines can be challenging, especially for users with multiple prescriptions or treatment plans. The system automates this process by extracting relevant details from prescriptions or user inputs and setting personalized reminders. 
	
	\textbf{Daily Health Report:}
	The system generates a comprehensive daily report to enhance health tracking, summarizing the user's essential health metrics and any detected anomalies throughout the day. This report is provided to the user each evening, presenting a clear overview of their health status. Furthermore, the reports are stored long-term, allowing users and healthcare providers to review historical data for well-informed decisions and trend analysis.

	\section{System Architecture}
	\label{sec:sys-arch}
	
	\begin{figure*}
		\begin{subfigure}{.5\linewidth}
			\centering
			\includegraphics[scale=0.5, width=0.9\linewidth]{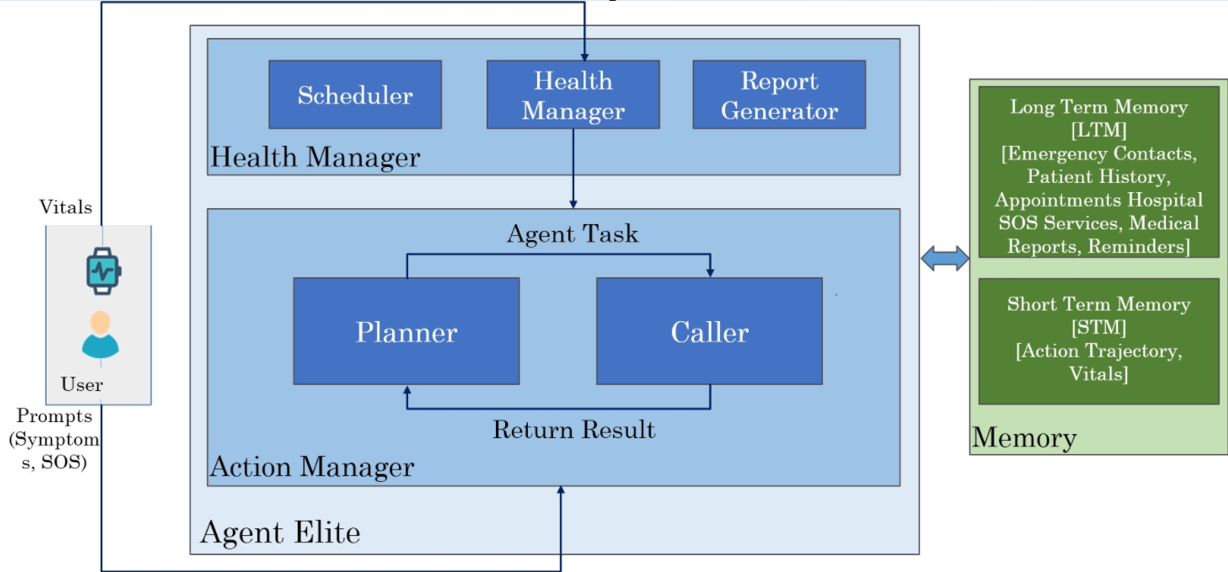}
			\caption{Communication in the multi-agent system}
			\label{fig:updatedArch}
		\end{subfigure}%
		\begin{subfigure}{.5\linewidth}
			\centering
			\includegraphics[scale=.5,width=0.9\linewidth]{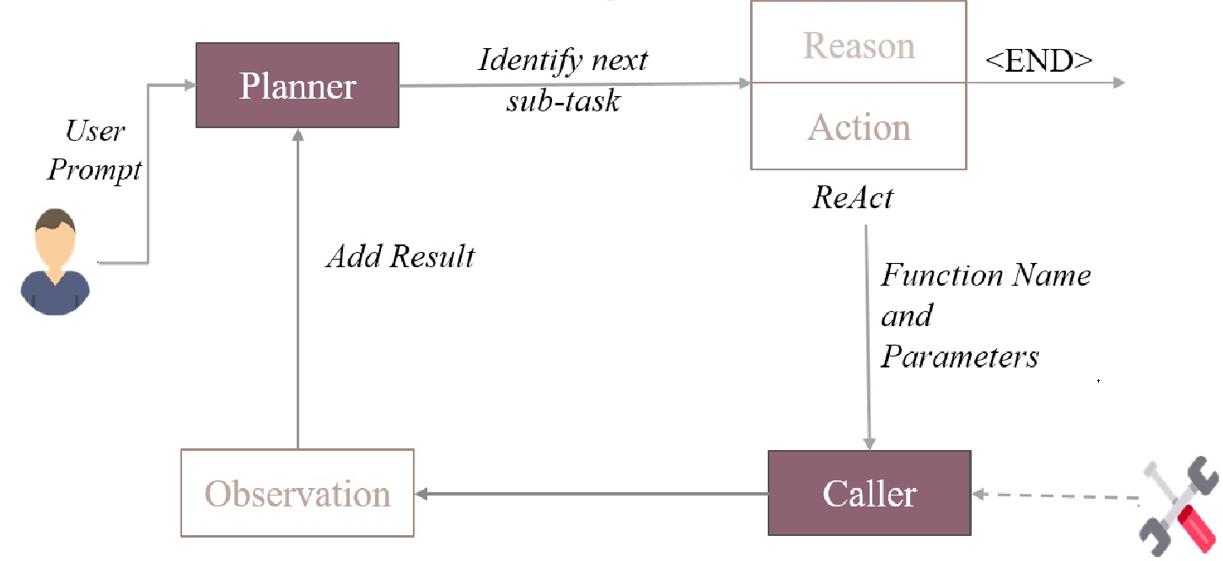}
			\caption{Communication between the planner and caller in Action Manager}
			\label{fig:actionManager}
		\end{subfigure}
		\caption{Multi-Agent Design for Healthcare Assistant}
	\end{figure*}
	In our proposed architecture, three key components collaborate to deliver a comprehensive healthcare assistance experience: the Action Manager, Health Manager, and Memory Unit. Figure \ref{fig:updatedArch} show the interaction between these three modules. The Health Manager's agents independently handle scheduling reminders, monitoring vitals, and generating reports. It connects with the Action Manager for user interaction and tool execution, utilizing the Action Manager's agents for planning and execution. The Memory Unit enables user-agent and agent-agent interaction, personalizing the system.
	We chose the Qwen 2.5-Coder-7B-Instruct model \footnote{Qwen2.5-Coder-7B-Instruct: \href{https://huggingface.co/Qwen/Qwen2.5-Coder-7B-Instruct}{https://huggingface.co/Qwen/Qwen2.5-Coder-7B-Instruct}} as the backbone model for our architecture due to its size which can fit in a flash memory and performance\cite{alizadeh2023llm}.	
	The following subsections delve into the details of each component.
	\subsection{Health Manager}
	The Health Manager is a crucial component of our multi-agent healthcare system that uses multiple  agents to monitor user health, generate reports, and manage medication schedules. It consists of three key sub-components: the Report Generator, Health Monitor, and Scheduler.
	
	\subsubsection{Report Generator}
	The Report Generator creates a comprehensive summary of each session using Qwen7B. It processes session data, including user queries, executed actions, vital health data, medicine prescriptions, and key insights gathered during the session, to generate a detailed and coherent report. This ensures that the session's key details are accurately captured and presented in a user-friendly format, enhancing the overall effectiveness and reliability of the system.
	
	\subsubsection{Health Monitor}
	The Health Monitor continuously analyzes vital signs from the user's wearable devices, such as smartwatches or smart rings, which provide real-time data on key health metrics like heart rate, blood pressure, and oxygen saturation. If the vitals deviate significantly from normal ranges, the system automatically triggers a soft SOS alert, serving as an early warning mechanism and ensuring that appropriate measures can be taken promptly.
	
	\subsubsection{Scheduler}
	The Scheduler assists users in managing their medication schedules effectively by analyzing the user's prescription and automatically extracting essential details such as medicine names, dosage timing, frequency, and duration. Using this information, the Scheduler creates personalized reminders that are sent to the user, helping them adhere to their prescribed treatment plan.
	\subsection{Action Manager}
	The role of Action Manager is to execute a task using available tools. This task is accomplished by two agent models, planner and caller. Since a model trained on code has function knowledge and improved reasoning \cite{ma2023training,zhang2024unveilingimpactcodingdata}, we choose an instruct finetuned model which has been also finetuned on coding tasks. We LoRA finetune it for planning and calling to incorporate planning and function calling abilities into the model and align with our usecases. 
	\subsubsection{Planner}
	The planner takes the user's query and based on the information of the tools, it generates an action along with its reasoning (ReAct framework \cite{yao2022react}). After the action is executed, the result along with the trajectory is passed to the planner, and the subsequent action and its reasoning is generated. Based on the results of the actions, next action is generated until the task is completed. This style of planning is called \underline{interleaved  planning} (\cite{huang2024understandingplanningllmagents}). We chose this over generating plan in a single-go for robustness in task execution.
	We LoRA finetune the Instruct-LLM model using our synthetic planning data (refer section \ref{sec:datagen}).
	
	\subsubsection{Caller}
	The Caller is responsible for identifying the correct parameters to be passed to the tool call. Our experiments revealed that function calling models perform poorly when dealing with a large number of tools (examples in section \ref{app-function-calling}). The Caller acts as the execution engine for various tasks, dynamically invoking appropriate functions based on instructions generated by the Planner \cite{zhang2024xlamfamilylargeaction,schick2023toolformerlanguagemodelsteach}. These tasks may include interacting with external APIs, retrieving or updating user-specific data, engaging with the user, and performing actions such as scheduling reminders, sending alerts, or interfacing with wearable devices.
	\subsection{Memory}
	The Memory Component is a crucial element in our multi-agent healthcare system, enabling a personalized user experience by storing and managing both long-term and short-term information. This component allows the system to tailor its interactions and provide more accurate, context-aware assistance \cite{sumers2024cognitivearchitectureslanguageagents, zhang2024surveymemorymechanismlarge}.
	
	\subsubsection{Short-Term Memory}
	The Short-Term Memory (STM) maintains context and temporary data during active sessions, such as the user's current symptoms and ongoing interactions. This enables the agents to maintain a coherent conversation and adapt to the user's immediate needs, ensuring a seamless and responsive experience.
	
	\subsubsection{Long-Term Memory}
	The Long-Term Memory (LTM) stores the user's personal information, previous symptoms, and other relevant data that can be retrieved by the Planner agent for improved decision-making. By leveraging this historical information, the Planner can better identify underlying issues and make more informed decisions when generating plans for the user's healthcare needs \cite{jiang2024longtermmemoryfoundation}. For example, knowing a patient's past complaints and vital history can help the Planner refer to the correct specialist, resulting in more targeted and effective care.
	
	For retrieving relevant information from the Long-Term Memory (LTM), we employ the retrieval model from the Spacy library's en\_core\_web\_trf \footnote{Spacy's  \href{https://github.com/explosion/spacy-models/releases/tag/en_core_web_trf-3.8.0}{en\_core\_web\_trf}} to extract pertinent data based on user queries and context. To parse the retrieved information, we utilize the EasyOCR library\footnote{EasyOCR:  \href{https://github.com/JaidedAI/EasyOCR}{https://github.com/JaidedAI/EasyOCR}}, which enables text recognition and extraction from various documents such as prescription.
	
	\section{Application Design}
	\label{sec:app-design}
	\begin{figure}[htbp]
		\includegraphics[width=\columnwidth]{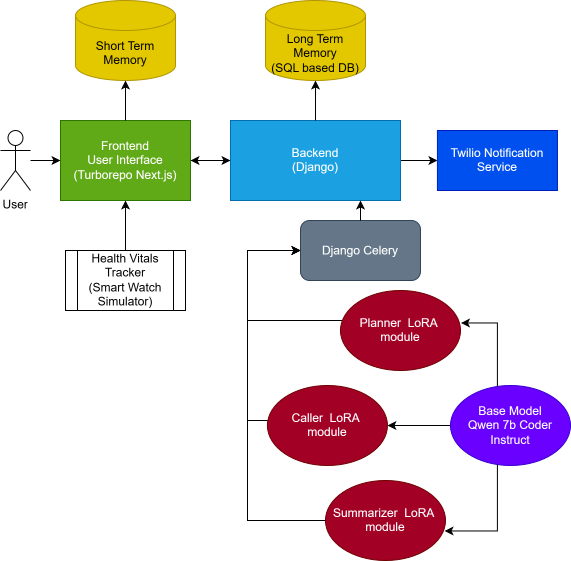}
		\caption{System flow diagram of the E2E application}
		\label{fig:app-arch}
	\end{figure}
	Our application follows a three-tier architecture, as illustrated in Figure \ref{fig:app-arch}. The user interacts with the system through the frontend, which serves as the topmost layer of the application. All user actions are processed at the backend, where appropriate responses are generated based on the specific action. The data layer, which consists of a SQLite database, handles data storage and retrieval for various queries and functionalities.
	
	Following are the main components of app architecture:
	\begin{itemize}
		\item Frontend – provides UI interface to the user for app interaction
		\item Backend – Intermediate processor between application services and frontend
		\item SQLite – SQL db to store necessary data
		\item Celery – Task queue to process async tasks
		\item Twilio – Sending messages to ambulance and emergency contacts
	\end{itemize}
	
	\subsection{Backend}
	The Django backend serves as the core of the application, handling various tasks such as processing user queries, managing user data, creating reminders based on uploaded prescriptions, triggering SOS calls, and interacting with the agent models. User interactions and histories are stored in a SQLite database, with Django ORM facilitating seamless data operations.
	
	The backend provides several key endpoints, including:
	\begin{itemize}
		\item Receiving user input for the agent model
		\item Presenting a summary of tracked vitals and suggestions
		\item Triggering SOS calls in case of abnormality detection
		\item Reminding users to take medication based on their prescription
	\end{itemize}
	
	The primary responsibilities of the backend include:
	\begin{itemize}
		\item Hosting multiple large action models
		\item Managing API endpoints for model interactions
		\item Providing a secure and scalable backend for the chat system and managing each user's history
		\item Analyzing recorded user vitals and providing suggestions based on the analysis
		\item Sending SMS to friends and ambulance services based on the severity of the detected abnormality
	\end{itemize}
	
	\subsection{Frontend}
	The frontend leverages Next.js\footnote{Next.js - \url{https://nextjs.org/}} for dynamic and responsive user interactions, managed under a Turborepo structure to streamline development across teams. Next.js, a React-based framework provides server-side rendering and client-side interactions for optimal user experience and at the same time provides room for scalability and improvement with well defined structure. Turborepo is used to manage multiple repositories under a single structure, optimizing builds and deployments, reducing development overhead for future.
	
	\subsubsection{UI Interface}
	The visual components are styled using tailwind CSS and styled-components, which provides pre-defined styling framework for fast as well as responsive development. 
	
	The functional components present in the UI are as follows:
	\begin{itemize}
		\item A real-time chat-like setup for users to interact with the planner model.
		\item	Powered by httpx and websockets for live updates and responsiveness.
		\item	Utilizes Fetch API to call Django REST endpoints.
		\item	Error handling and loading states ensure a smooth user experience.
		\item	Watch and SOS simulator have been included for demo purpose.
		\item	Agent logs are added to help understand the logic behind agent’s inference for user input.
	\end{itemize}
	
	\subsubsection{State Management}
	\begin{itemize}
		\item 	State is managed using React Context API for lightweight and efficient state sharing.
		\item 	Server-side rendering (SSR) for initial page load and client-side fetching for subsequent interactions to save on data and improve response time.
	\end{itemize}
	
	\section{Data Creation}
	\label{sec:datagen}
	
	We curate a synthetic dataset to LoRA finetune the planner and caller (refer table \ref{tab:dataset_expansion}). We present our data generation process in the following subsections.
	\begin{table}[h]
		\centering
		\begin{tblr}{
				cell{3}{1} = {r=4}{},
				vlines,
				hline{1,10} = {-}{0.08em},
				hline{2-3} = {-}{},
				hline{4-9} = {2-4}{},
			}
			\textbf{Task} & \textbf{Usecase} & \textbf{Planner} & \textbf{Caller} \\
			Train         & Combined         & 374,000          & 328,000         \\
			Test          & General          & 700              & 800             \\
			& Counter          & 876              & 976             \\
			& Negative         & 492              & 589             \\
			& Dietician        & 1,273             & 1,423            \\
			& Soft SOS & 200 & 100 \\
			& Hard SOS - Start & 700 & 600 \\
			& Hard SOS - End & 600 & 500 
		\end{tblr}
		\caption{Statistics of Synthetic Data Created for Planner and Caller agents. Combined indicates mixture of appointment usecases (general, counter, negative, dietician), hard SOS (start and end) and soft SOS. General: Appointment booked after finding specialist for the symptoms, Counter: Follow-up questions asked to understand symptoms better, Negative: User declines the appointment, Dietician: Use is referred to dietician based on the symptoms} 
		\label{tab:dataset_expansion}
	\end{table}
	\begin{figure}[htbp]
		\centering
		\includegraphics[width=\columnwidth]{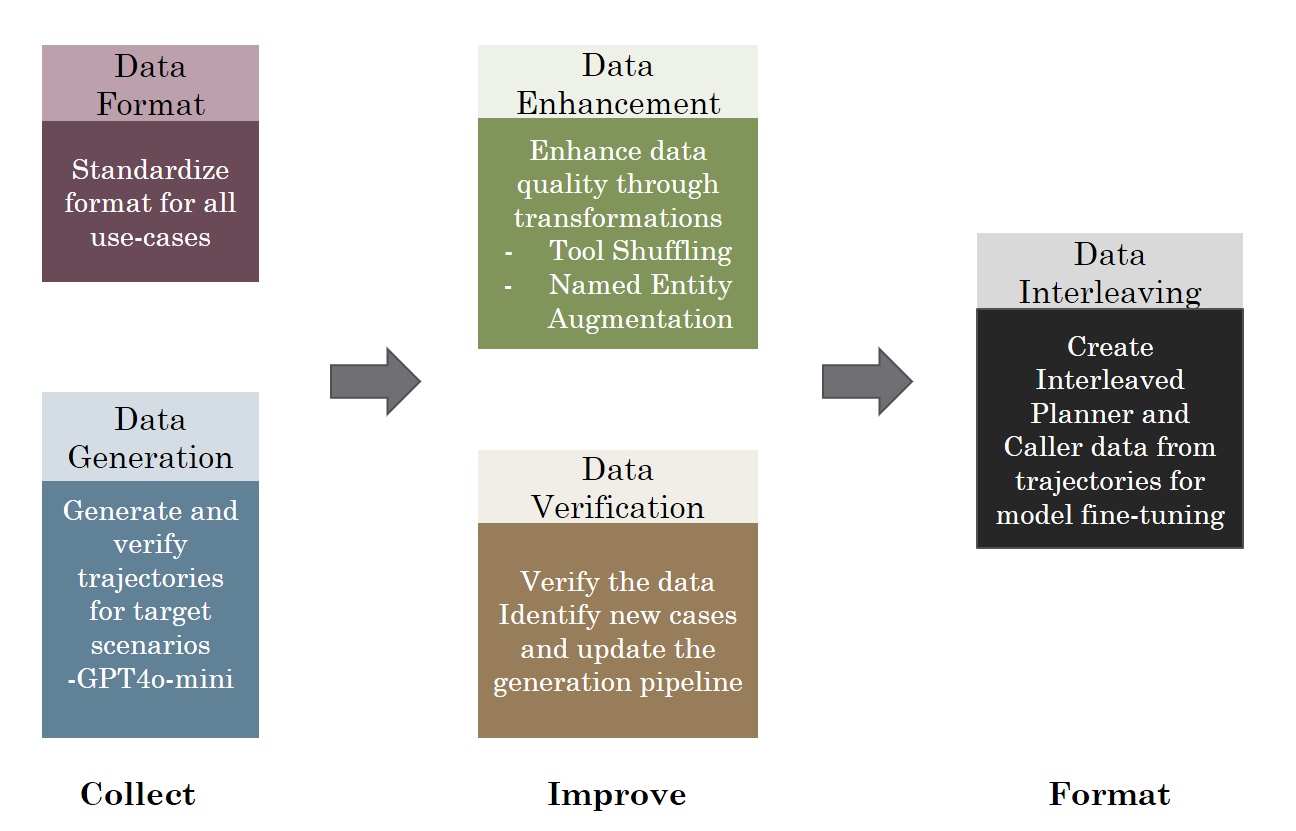}
		\caption{Data Creation Process}
		\label{fig:data_gen}
	\end{figure}
	
	\subsection{Data Gathering}
	We first begin by standardizing the format of the data. We create a trajectory containing communication between the planner and caller to complete the task. The planner uses the ReAct framework where based on the task the next action and its reasoning is generated. The caller generates the function call along with appropriate parameters. After finalizing format of trajectory, we generate data using OpenAI's gpt-4o-mini.
	\subsubsection{Data Format}
	This section introduces the data format used in our multi-agent healthcare system. A trajectory represents an interleaved history of the interaction between the user, multiple agents, and the environment to complete a task.
	Following is the format of each state in the trajectory:
	\begin{enumerate}
		\item System
		\begin{lstlisting}
			{
				"from": "system",
				"value": {
					"user_details": {
						"user_id": "BKYA569367",
						"name": "John Doe",
						"timestamp": "2024-12-17T23:59:59"
					} } }
		\end{lstlisting}
		\item User
		\begin{lstlisting}
			{
				"from": "user",
				"value": "Natural language input from user"
			}
		\end{lstlisting}
		\item Planner
		\begin{lstlisting}
			{
				"from": "planner",
				"value": {
					"reason": "Reasoning Generated by LLM based on previous actions",
					"action": "Next action based on LLM's reasoning. <END> for completed task"
				} }
		\end{lstlisting}
		\item Caller
		\begin{lstlisting}
			{
				"from": "caller",
				"value": {
					"tool": "function_name",
					"parameters": {
						"param1": "value1",
						"param2": "value2",
						...
						"paramN": "valueN"
					} } }
		\end{lstlisting}
		\item Observation
		\begin{lstlisting}
			{
				"from": "observation",
				"value": {
					"result": Object_returned_by_function
				} }
		\end{lstlisting}
		Consolidating the states, the format of trajectory looks as follows:
		\begin{lstlisting}
			{
				"from": "system",
				"value": {...}
			},
			{
				"from": "user",
				"value": ""
			},
			{
				"from": "planner",
				"value": {...}
			},
			{
				"from": "caller",
				"value": {...}
			},
			{
				"from": "observation",
				"value": {...}
			}
			...
			{
				"from": "planner",
				"value": {
					"reason": "The task is sucessfully completed.",
					"action": "<END>"
				} }
		\end{lstlisting}
		For our usecases, examples of trajectories are shown in section \ref{app-trajectory}.
	\end{enumerate}
	\subsubsection{Data Generation}
	To generate data for our multi-agent healthcare system, we use a 1-shot prompt with OpenAI's gpt-4o-mini, providing a list of tools, a description of the use case, and an example trajectory. For instance, when generating appointment data, we want the Planner to suggest a specialist without diagnosing a disease while maintaining data diversity. To achieve this, we include a specific disease in the prompt and instruct the LLM to generate symptoms for that disease and suggest a specialist accordingly.
	\subsection{Data Enhancement}
	After generating the data, we noticed that it had repetitive names, dates, and times. To fix this, we replaced them with random values. Since the input and output formats of tools for the named entities are standard (e.g. YYYY-MM-DD for date), we used regular expressions to find these entities. For dates and times, we made sure that in cases like appointments, the date of past symptoms comes before the user's query date, and the suggested appointment is within a reasonable time after the user's query.
	
	The tools and their descriptions in the input to the Planner serve as premises for reasoning and generating the next action. As demonstrated by \cite{chen2024premise}, shuffling the premises can hinder the model's reasoning ability. Therefore, we shuffle the tools for each sample to enhance the model's robustness and adaptability..
	\subsection{Data Verification}
	We analyze the generated data for each usecase and identify the scope of improvements. Apart from data augmentation, we check cases where the trained models might fail with the current data and adapt it for such exigencies. For example, when user gives a few symptoms and planner might select an incorrect specialist for appointment. To tackle this, we allow the planner to counter questions and prompt gpt-4o-mini to generate data where user inputs vague symptoms at first.
	\subsection{Data Interleaving}
	The trajectory consists of a sequence of states that include system, user, planner, and caller interactions. To create datasets for the Planner and Caller agents, we employ data interleaving. For each Planner state in the trajectory, we select the tools, previous interactions, and observations from function calls as input, while the reason and action are chosen as output. Similarly, for each Caller state, the function call and its parameters are set as output.
	
	In simpler terms, we extract relevant information from the trajectory to create training data for the Planner and Caller agents. This process involves identifying the appropriate input and output for each agent based on their specific roles within the sequence of states. By creating these datasets, we enable the fine-tuning of the Planner and Caller models to improve their performance in generating accurate reasoning, actions, and function calls.
	
	We enclose the reason, action, function and parameters in tags for convenient post-processing. We also add these as special tokens in the model. For example, the output of planner for booking a specialist for the user:
	
	\begin{lstlisting}
		<reason>Dr. Gabriel Lopez (Neurologist) is available on the user's preferred date and time.</reason>
		<action>Suggest the appointment to the user and proceed with booking if confirmed. Dr. Gabriel Lopez (Neurologist) on 2024-12-01 between 10:00-10:30</action>
	\end{lstlisting}
	\begin{table*}[h]
		\centering
		\begin{tabular}{|l|c|c|c|c|}
			\hline
			\textbf{Category} & \textbf{BLEU} & \textbf{Rouge1} & \textbf{Rouge2} & \textbf{RougeL} \\ \hline
			General  & 94.79 & 0.97 & 0.95 & 0.97 \\ \hline
			Counter  & 78.97 & 0.84 & 0.76 & 0.83 \\ \hline
			Negative & 78.05 & 0.86 & 0.76 & 0.84 \\ \hline
			Dietician  & 74.44 & 0.80 & 0.71 & 0.78 \\ \hline
		\end{tabular}
		\caption{Evaluation results of Planner for appointment usecase. General: Appointment booked after finding specialist for the symptoms, Counter: Follow-up questions asked to understand symptoms better, Negative: User declines the appointment, Dietician: Use is referred to dietician based on the symptoms}
		\label{tab:results-planner}
	\end{table*}
	\begin{table*}[h]
		\centering
		\begin{tabular}{|l|c|c|c|c|c|c|c|c|}
			\hline
			\textbf{Category}  & \textbf{BLEU} & \textbf{Rouge1} & \textbf{Rouge2} & \textbf{RougeL} & \textbf{ToolAcc} & \textbf{ParamAcc} & \textbf{ValuesAcc} \\ \hline
			General & 98.88 & 1.0 & 0.99 & 0.99 & 1.0 & 1.0 & 0.9 \\ \hline
			Counter & 96.95 & 0.98 & 0.96 & 0.98 & 1.0 & 0.99 & 0.77 \\ \hline
			Negative & 89.64 & 0.92 & 0.88 & 0.92 & 1.0 & 0.93 & 0.57 \\ \hline
			Dietician & 96.06 & 0.98 & 0.95 & 0.97 & 1.0 & 1.0 & 0.76 \\ \hline
		\end{tabular}
		\caption{Evaluation results of Caller for appointment usecase. General: Appointment booked after finding specialist for the symptoms, Counter: Follow-up questions asked to understand symptoms better, Negative: User declines the appointment, Dietician: Use is referred to dietician based on the symptoms}
		\label{tab:results-caller}
	\end{table*}
\section{Experimental Setup and Results}
\label{sec:results}
	We fine-tuned the Qwen2.5-Coder-7B-Instruct model for our multi-agent healthcare assistant using synthetic data and LoRA adaptation with rank 8 and alpha 16. Context length was set to 4,096 tokens. The model was trained for 5,000 steps with a batch size of 64. By fine-tuning on separate Planner and Caller datasets, we achieved a modular system for edge device deployment, with each LoRA adapter taking only 10MB before compression.
		
	We evaluate the performance of our multi-agent healthcare system using distinct metrics for the Planner and Caller agents. For the Planner, we assess the correctness of its reasoning and action using BLEU and ROUGE scores. ROUGE and BLEU scores are apt for checking the correctness of the Planner's reasoning and action as they quantitatively assess the similarity between the generated text and reference text, providing a standardized way to evaluate the quality of the Planner's responses. In the case of the Caller, we measure the accuracy of the tool called (ToolAcc), parameters passed to the tool (ParamAcc) and values of the parameters (ValuesAcc). We also employ BLEU and ROUGE scores for the Caller, as they prove valuable when a natural language sentence is provided as input to the tool.
	
	On our test set, we achieve an average RougeL score of 85.5 for planner and 96.5 for caller for the appointment usecase. Notably, the Caller consistently predicts the correct tool. For detailed scores for appointment usecase, refer to Table \ref{tab:results-planner} and Table \ref{tab:results-caller}. For SOS usecases, we observe 100\% scores for both planner and caller because of a fixed sequence of actions.
\section{Conclusion}
	Our multi-agent healthcare system demonstrates the potential for AI-powered assistants to revolutionize personal health management. By leveraging the power of large language models and fine-tuning them for specific tasks, we have created a modular and efficient system that can effectively support users in managing their health and well-being.
	
	The Action Manager, Health Manager, and Memory Unit work together to provide a personalized healthcare experience. The Action Manager coordinates tasks, the Health Manager monitors health and manages schedules, while the Memory Unit stores and retrieves user-specific information for context-aware assistance.
	
	Through extensive data generation, enhancement, and fine-tuning, we have achieved a RougeL scores of 85.5 for planning and 96.5 for calling. This demonstrates the system's ability to generate relevant and accurate responses, while the Caller's consistent prediction of the correct tool showcases its reliability in executing tasks.
\section{Future Work}
	We plan to enhance our multi-agent system by converting the trained modules into on-device compatible formats like ONNX/GGUF, enabling smooth system integration into mobile devices and lower latency. We'll also integrate the system with various health monitoring devices and provide support for user onboarding of additional devices like smart rings and smart weighing scales. To enhance healthcare capabilities, we'll extend the system into a multimodal understanding engine, empowering it to perform image-based diagnosis. Furthermore, we aim to expand the system to cover additional targeted scenarios, such as women's health tracking and post hospitalization care.

	\bibliography{custom}
	
	\appendix
	\section{Snapshots of Application}
	In this section, we present snapshots showcasing the key features of our application. These include intelligent appointment scheduling based on user symptoms, reminder scheduling by extracting prescription information, soft SOS triggers for abnormal vitals, and user-initiated hard SOS with location tracking and emergency notifications.
	\subsection{Intelligent Appointment Scheduling}
\begin{figure*}
		\centering
		\includegraphics[scale=0.2]{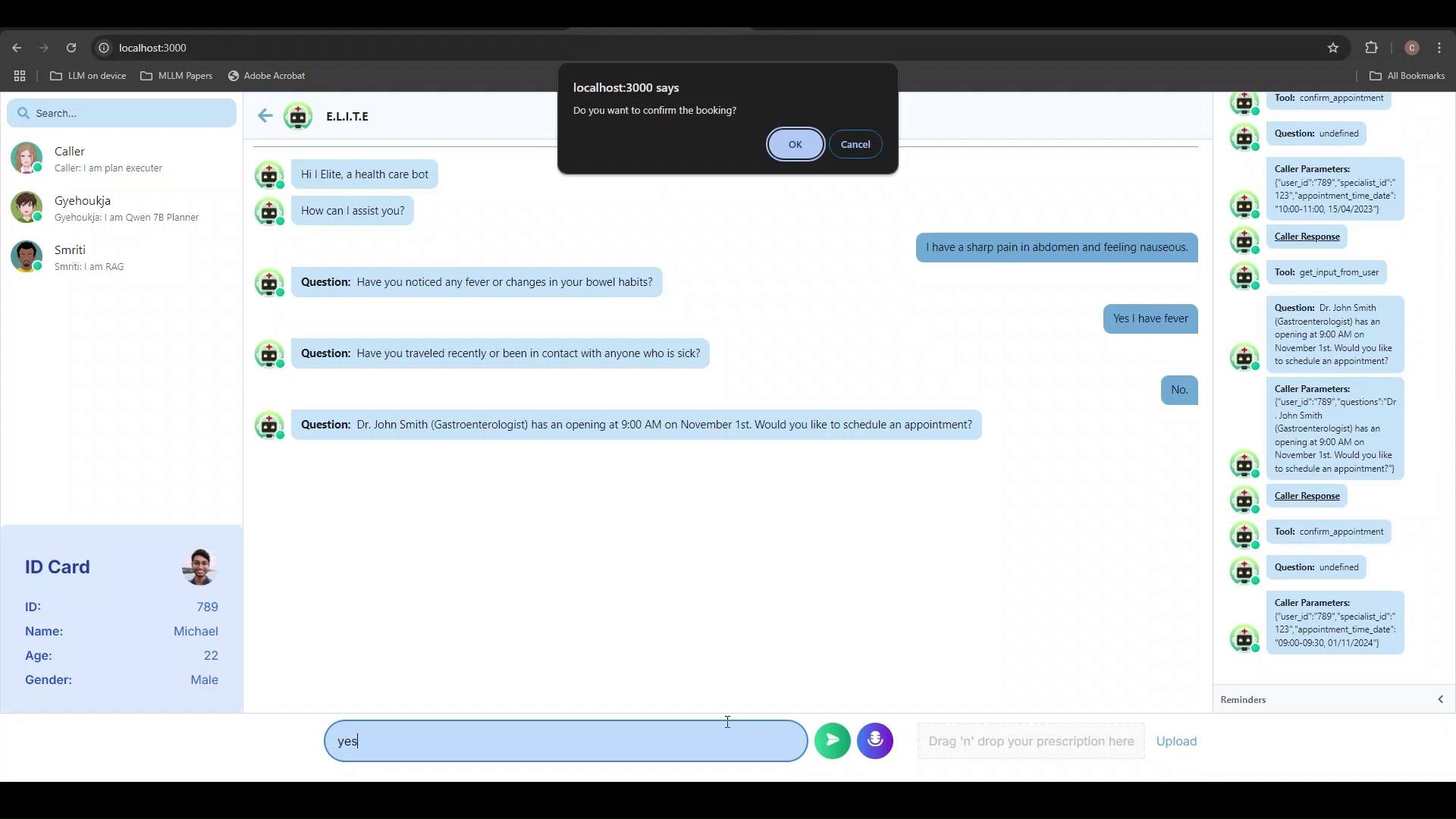}
		\caption{Appointment Booking}
		\label{fig:usecase-appointment}
	\end{figure*}

	Figure \ref{fig:usecase-appointment} shows the appointment booking use case where the user complains about having abdominal pain and feeling feverish. The Planner and Caller ask follow-up questions and find a specialist that fits the user's schedule.
	\subsection{Reminder Scheduling from Prescription}
\begin{figure*}
		\begin{subfigure}{.5\linewidth}
			\centering
			\includegraphics[scale=0.5, width=0.9\linewidth]{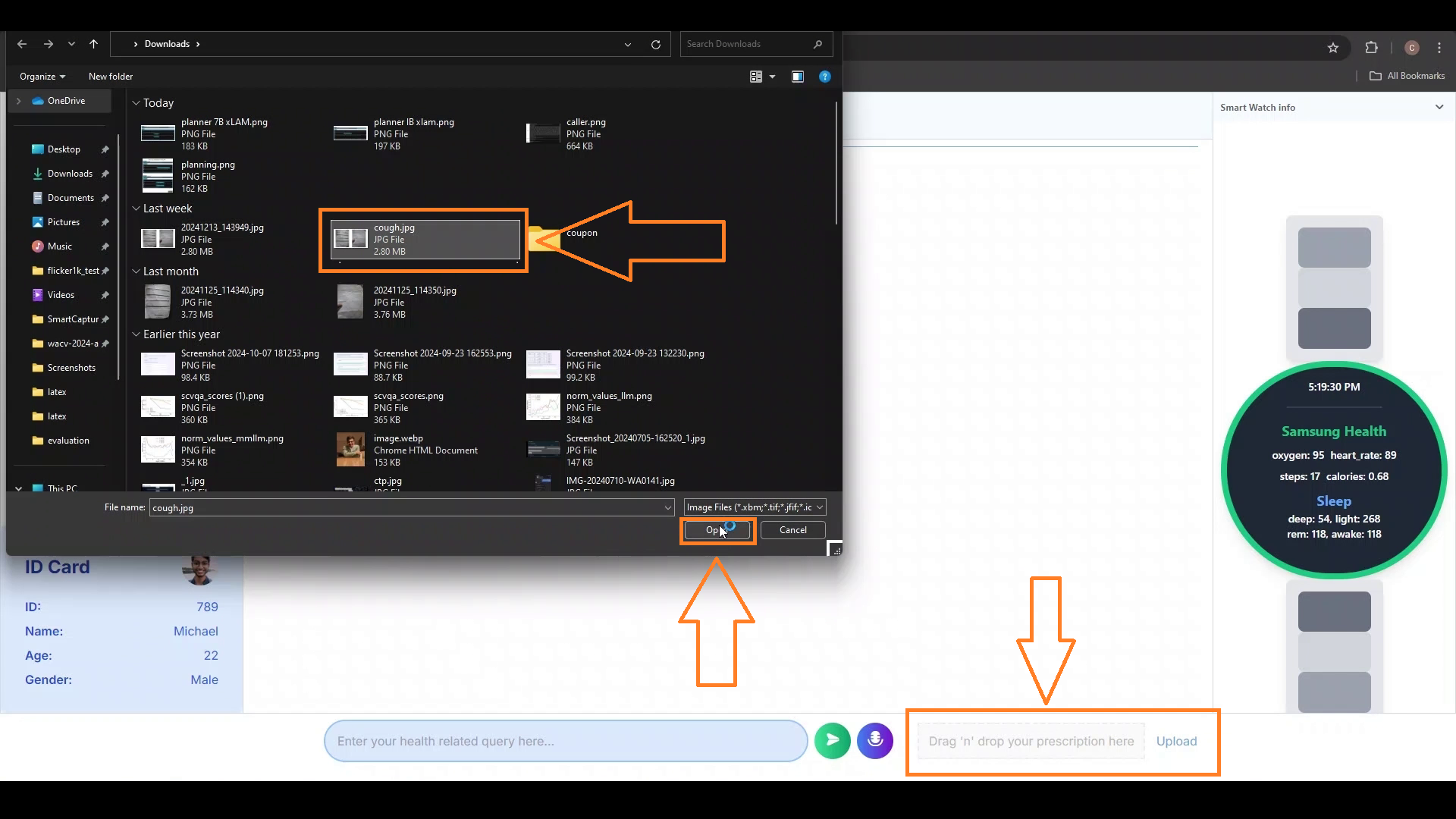}
			\caption{Medication Uploaded schedule Reminders}
			\label{fig:reminder-upload}
		\end{subfigure}%
		\begin{subfigure}{.5\linewidth}
			\centering
			\includegraphics[scale=0.5,width=0.9\linewidth]{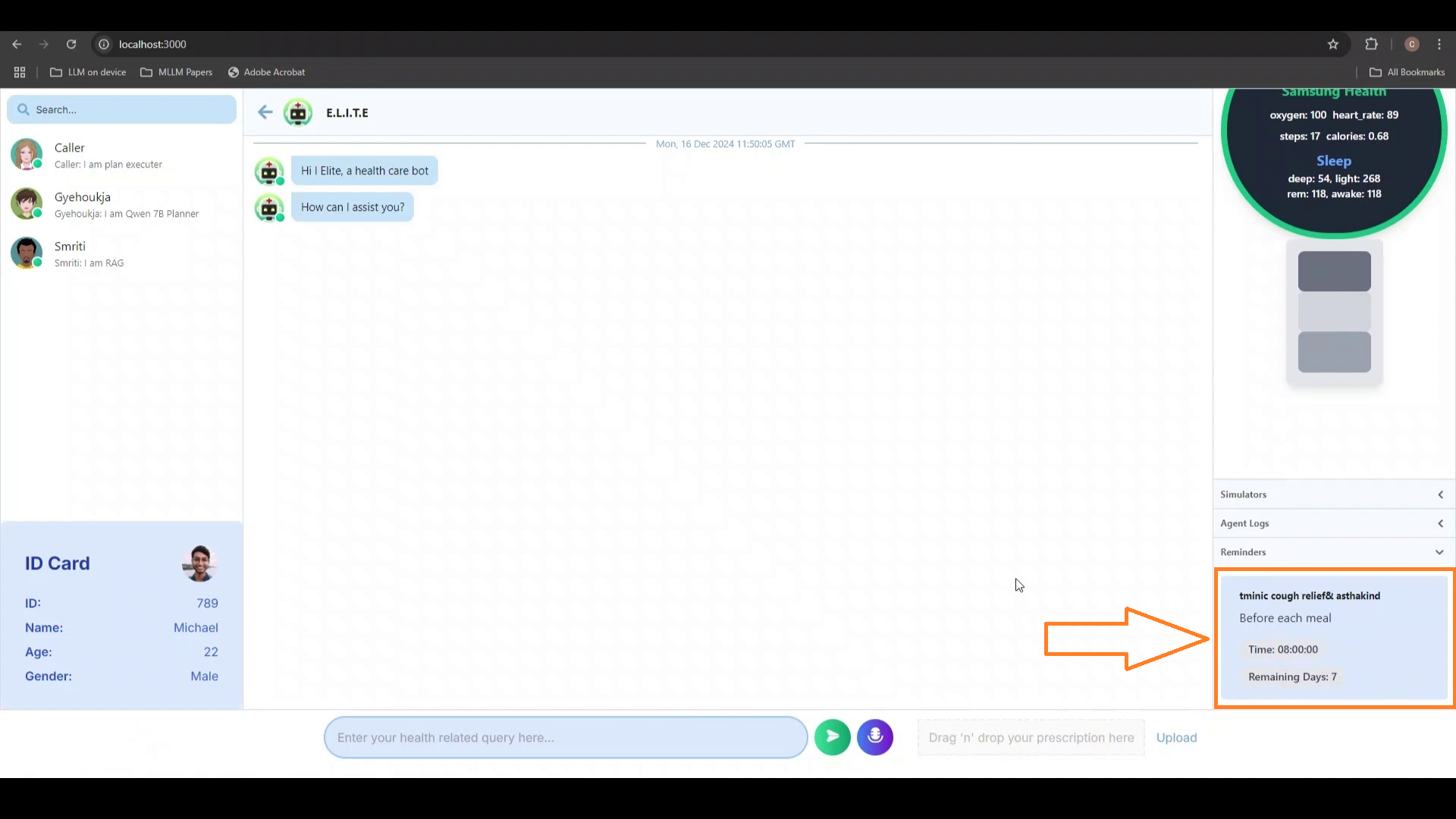}
			\caption{Reminder Added}
			\label{fig:usecase-reminder-set}
		\end{subfigure}
		\caption{Adding Reminder from Prescription}
		\label{fig:usecase-reminder}
	\end{figure*}

	As shown in figure \ref{fig:reminder-upload}, the user can upload a prescription. The prescription is parsed and analyzed by LLM to extract information about medication such as medicine name, time, and duration. This information is then added to the user's list of reminders, as shown in figure \ref{fig:usecase-reminder-set}.
	\subsection{Soft Trigger from Abnormal Vitals}
\begin{figure*}
			\begin{subfigure}{.5\linewidth}
				\centering
				\includegraphics[scale=0.5, width=0.9\linewidth]{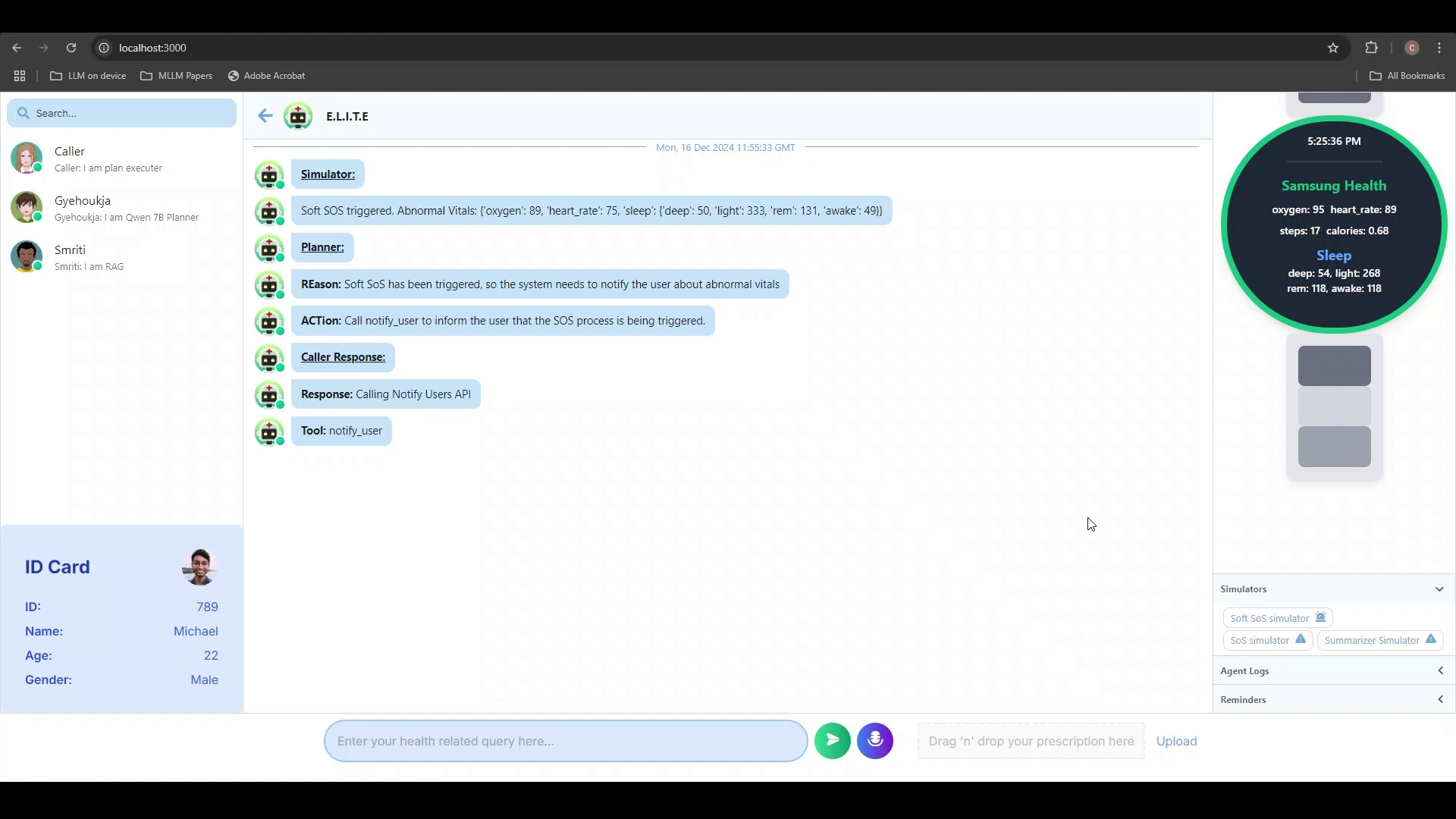}
				\caption{Soft SOS triggered due to abnormal vitals (simulated)}
				\label{fig:soft-sos-triggered}
			\end{subfigure}%
			\begin{subfigure}{.5\linewidth}
				\centering
				\includegraphics[scale=0.5,width=0.9\linewidth]{reminder-set.png}
				\caption{User's vitals analyzed}
				\label{fig:usecase-softsos}
			\end{subfigure}
			\caption{Soft SOS triggered and vitals analyzed}
			\label{fig:usecase-softsos-and-analysis}
		\end{figure*}

	If a user's vitals are abnormal, the monitor agent triggers a soft SOS. At that time, the user's vitals are fetched (simulated for our demo) and a small report is displayed to the user (refer to figure \ref{fig:usecase-softsos-and-analysis}).
	\subsection{Hard SOS Triggered by User}
\begin{figure*}
			\begin{subfigure}{.5\linewidth}
				\centering
				\includegraphics[scale=0.22]{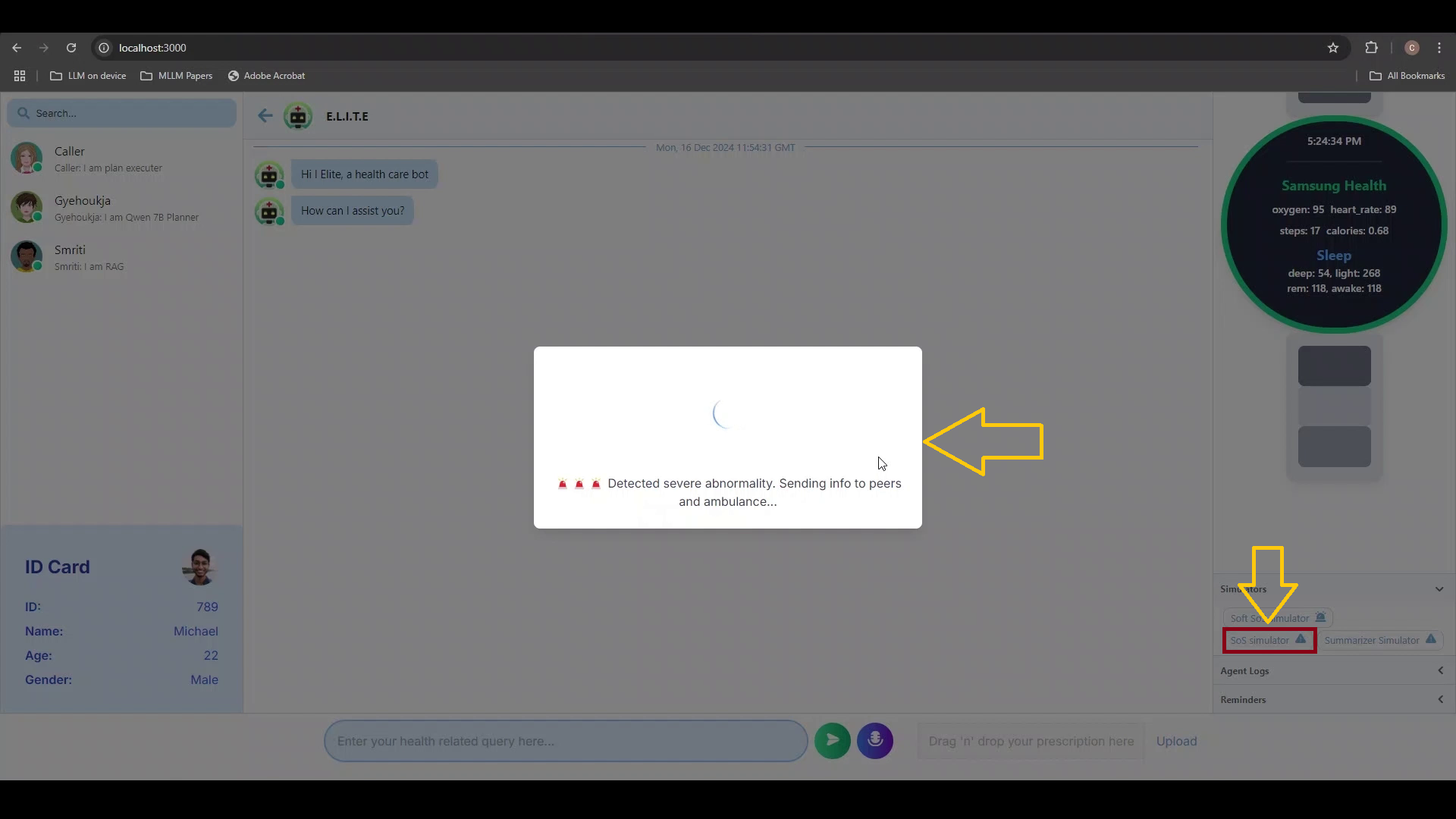}
				\caption{Hard SOS-Interface}
				\label{fig:usecase-hardsos}
			\end{subfigure}%
			\begin{subfigure}{.5\linewidth}
				\centering
				\includegraphics[scale=0.2]{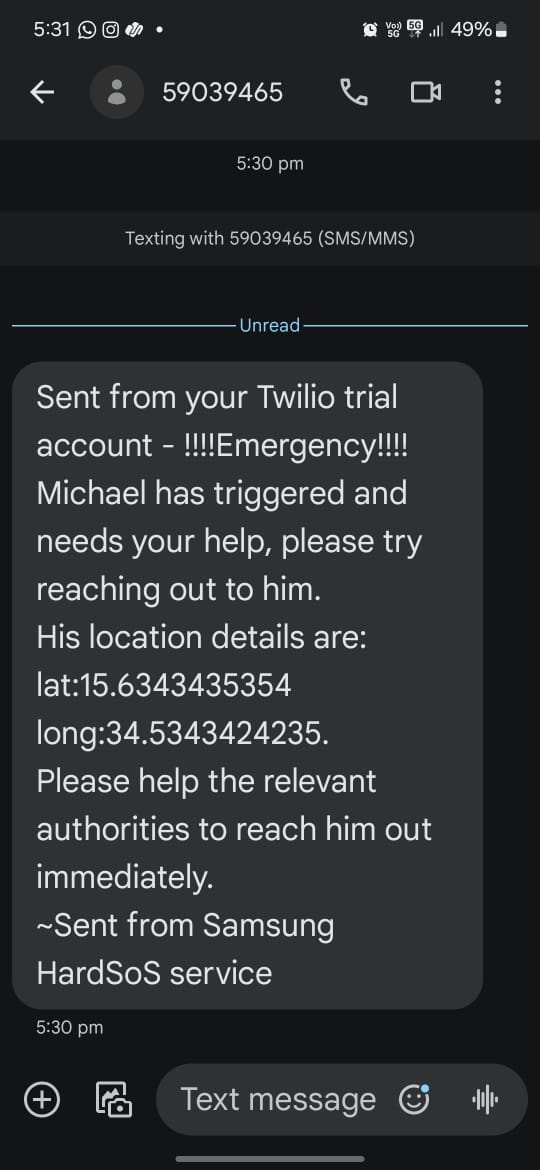}
				\caption{Hard SOS-SMS}
				\label{fig:usecase-sms}
			\end{subfigure}
			\caption{Triggering Hard SOS and notifying emergency contacts with GPS}
			\label{fig:usecase-hardsos-and-sms}
		\end{figure*}

	The user can trigger a Hard SOS as shown in figure \ref{fig:usecase-hardsos}. The user's location is fetched and the nearest ambulance is identified. The location is then sent to the nearest ambulance and emergency contacts. A sample of the message sent to an emergency contact is shown in figure \ref{fig:usecase-sms}.
	
		
		\begin{figure*}
			\centering
			\includegraphics[scale=0.5]{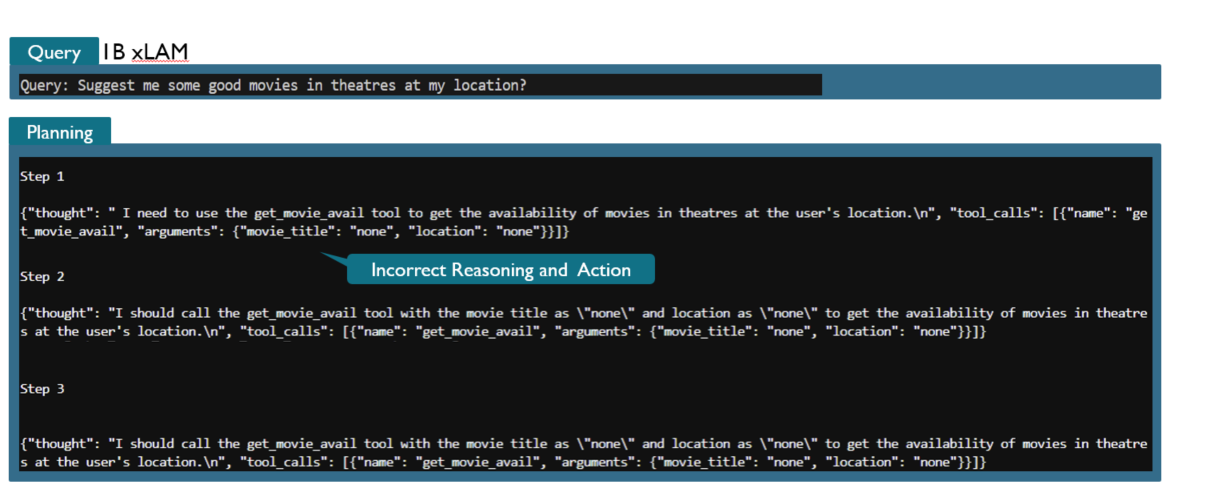}
			\caption{Planning and function calling using xLAM 1b}
			\label{fig:exp-plan1}
		\end{figure*}
		\begin{figure*}
			\centering
			\includegraphics[scale=0.5]{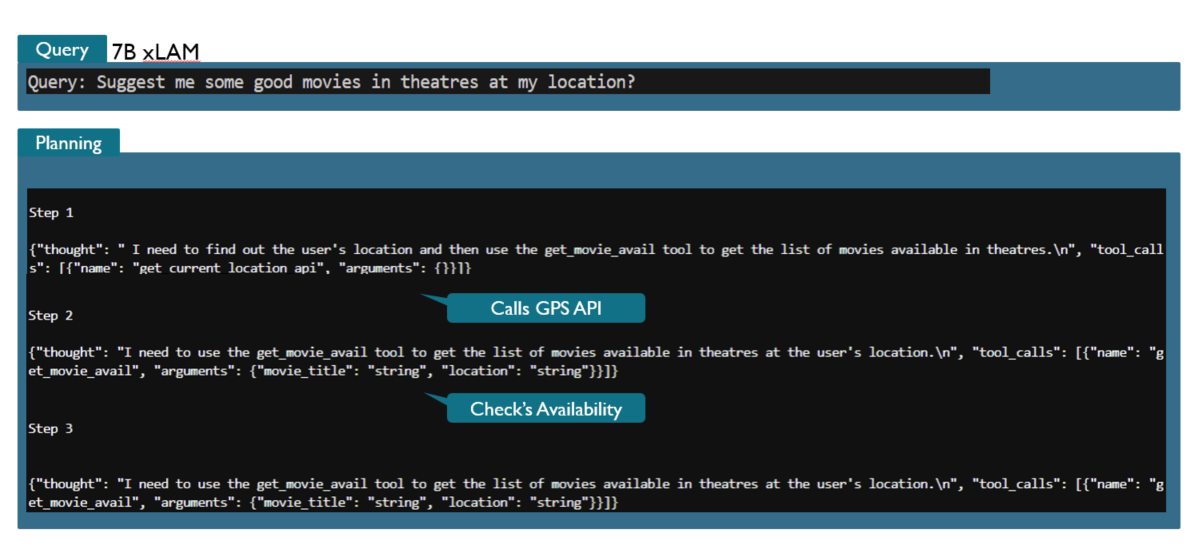}
			\caption{Planning and function calling using xLAM 7b}
			\label{fig:exp-plan7}
		\end{figure*}
	\begin{figure*}[htbp]
		\includegraphics[width=2\columnwidth]{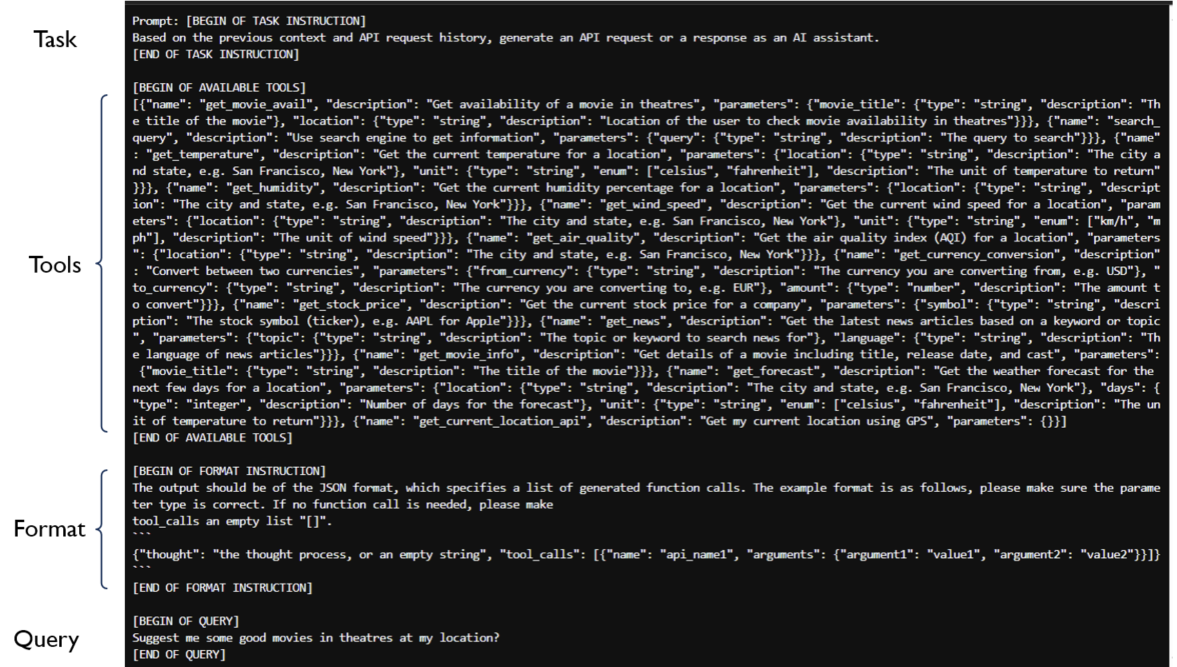}
		\caption{Prompt to xLAM 7b}
		\label{fig:exp-call}
	\end{figure*}

	\label{app-snaps-app}
	\section{Experimentation with function calling models}
	\label{app-function-calling}
	We conducted experiments with xLAM 1B and 7B models to assess their suitability for planning and calling tasks. As illustrated in Figure \ref{fig:exp-call}, we provided a list of 28 tools for the task of finding good movies in nearby theaters. However, both the 1B and 7B models failed to identify the correct function calls in steps 1 and 2, respectively (refer figure \ref{fig:exp-plan1} and \ref{fig:exp-plan7}). Specifically, the 7B model attempted to check movie availability before fetching the list of movies. To address these issues, we fine-tuned the Qwen model to improve its performance and avoid such situations.
	\section{List of Diseases}
	\label{app-diseases}
	During data generation, the prompt instructs to generate symptoms based on a given disease. We randomly sample from a list of 153 diseases. The list was created based on different specialization such as orthologist, cardiologist, gastroenterologists, etc. They are as follows: Influenza, Diabetes Mellitus, Hypertension, Asthma, Coronary Artery Disease, Stroke, HIV/AIDS, Tuberculosis, Hepatitis B, Hepatitis C, Dengue Fever, Malaria, Chickenpox, Measles, Mumps, Rubella, Zika Virus, Ebola, COVID-19, Cancer, Alzheimer's Disease, Parkinson's Disease, Arthritis, Lupus, Multiple Sclerosis, Crohn's Disease, Ulcerative Colitis, Gastroesophageal Reflux Disease (GERD), Celiac Disease, Anemia, Bronchitis, Pneumonia, Sinusitis, Chlamydia, Gonorrhea, Syphilis, Herpes, Human Papillomavirus (HPV), Tetanus, Rabies, Polio, Yellow Fever, Smallpox, Typhoid Fever, Leprosy, Plague, Hemophilia, Sickle Cell Disease, Thalassemia, Hemochromatosis, Meningitis, Encephalitis, Glaucoma, Cataracts, Macular Degeneration, Otitis Media, Tonsillitis, Appendicitis, Hemorrhoids, Diverticulitis, Irritable Bowel Syndrome (IBS), Kidney Stones, Chronic Kidney Disease, Renal Failure, Osteoporosis, Osteopenia, Rheumatoid Arthritis, Gout, Psoriasis, Eczema, Vitiligo, Alopecia, Hypothyroidism, Hyperthyroidism, Cushing's Syndrome, Addison's Disease, Prostate Cancer, Breast Cancer, Lung Cancer, Colorectal Cancer, Pancreatic Cancer, Liver Cancer, Skin Cancer, Melanoma, Leukemia, Lymphoma, Myeloma, Basal Cell Carcinoma, Squamous Cell Carcinoma, Esophageal Cancer, Gastric Cancer, Ovarian Cancer, Cervical Cancer, Endometrial Cancer, Bladder Cancer, Kidney Cancer, Testicular Cancer, Thyroid Cancer, Mesothelioma, Brain Tumors, Spinal Cord Tumors, Heart Failure, Atrial Fibrillation, Cardiomyopathy, Angina, Atherosclerosis, Pericarditis, Myocarditis, Endocarditis, Congenital Heart Disease, Valvular Heart Disease, Peripheral Artery Disease, Deep Vein Thrombosis, Pulmonary Embolism, Varicose Veins, Hypertensive Heart Disease, Coronary Microvascular Disease, Dyslipidemia, Familial Hypercholesterolemia, Hypertrophic Cardiomyopathy, Dilated Cardiomyopathy, Restrictive Cardiomyopathy, Takotsubo Cardiomyopathy, Sleep Apnea, Insomnia, Narcolepsy, Restless Leg Syndrome, Epilepsy, Migraine, Cluster Headache, Tension Headache, Trigeminal Neuralgia, Multiple System Atrophy, Amyotrophic Lateral Sclerosis (ALS), Huntington's Disease, Dystonia, Myasthenia Gravis, Guillain-Barre Syndrome, Spinal Muscular Atrophy, Muscular Dystrophy, Fibromyalgia, Chronic Fatigue Syndrome, Complex Regional Pain Syndrome, Temporomandibular Joint Disorders, Sjogren's Syndrome, Raynaud's Disease, Scleroderma, Mixed Connective Tisse Disease, Behcet's Disease, Wegener's Granulomatosis, Giant Cell Arteritis, Polyarteritis Nodosa and Sarcoidosis.
	\section{Example of Trajectory}
	\label{app-trajectory}
	\subsection{Appointment - Generic}
	\label{app-generic-appointment-traj}
	\begin{lstlisting}
		{
			"interaction_trajectory": [
			{
				"from": "system",
				"value": {
					"user_details": {
						"user_id": "JICC571413",
						"name": "Sakura Tominaga",
						"timestamp": "2024-09-02T10:57:00"
					}
				}
			},
			{
				"from": "user",
				"value": "I've been feeling extremely fatigued with chills, body aches, and a sore throat. It's becoming hard to get through the day."
			},
			{
				"from": "planner",
				"value": {
					"reason": "User reports fatigue, chills, body aches, and a sore throat which could indicate a viral infection such as influenza.",
					"action": "Check for past similar complaints using the retrieve_past_complaints tool."
				}
			},
			{
				"from": "caller",
				"value": {
					"tool": "retrieve_past_complaints",
					"parameters": {
						"user_id": "JICC571413",
						"symptoms": "fatigue, chills, body aches, sore throat"
					}
				}
			},
			{
				"from": "observation",
				"value": {
					"result": [
					{
						"date": "2024-06-02",
						"symptoms": "mild body aches, slight fever"
					}
					]
				}
			},
			{
				"from": "planner",
				"value": {
					"reason": "Past records indicate a mild common cold, but the current symptoms are more severe. A consultation with a general physician is advisable.",
					"action": "Use the get_available_specialists tool to find a general physician available for an appointment."
				}
			},
			{
				"from": "caller",
				"value": {
					"tool": "get_available_specialists",
					"parameters": {
						"symptoms": "fatigue, chills, body aches, sore throat",
						"specialization": "general physician"
					}
				}
			},
			{
				"from": "observation",
				"value": {
					"result": {
						"specialist_id": "AECJ317777",
						"name": "Dr. Diego Arroyo (General Physician)",
						"available_slot": {
							"date": "2024-11-30",
							"time": "11:00-11:30"
						}
					}
				}
			},
			{
				"from": "planner",
				"value": {
					"reason": "Dr. Diego Arroyo (General Physician) is available on the user's preferred date and time.",
					"action": "Suggest the appointment to the user and proceed with booking if confirmed. Dr. Diego Arroyo (General Physician) on 2024-11-30 between 11:00-11:30"
				}
			},
			{
				"from": "caller",
				"value": {
					"tool": "get_input_from_user",
					"parameters": {
						"user_id": "JICC571413",
						"questions": "Dr. Diego Arroyo (General Physician) has an opening at 11:00 AM on November 30th. Would you like to schedule an appointment?"
					}
				}
			},
			{
				"from": "observation",
				"value": {
					"result": "Yes, please"
				}
			},
			{
				"from": "planner",
				"value": {
					"reason": "User has confirmed the appointment",
					"action": "I should confirm the appointment"
				}
			},
			{
				"from": "caller",
				"value": {
					"tool": "confirm_appointment",
					"parameters": {
						"user_id": "JICC571413",
						"specialist_id": "AECJ317777",
						"appointment_time_date": "11:00-11:30, 30/11/2024"
					}
				}
			},
			{
				"from": "observation",
				"value": {
					"result": true
				}
			},
			{
				"from": "planner",
				"value": {
					"reason": "Appointment confirmed. Next, I should save the appointment history.",
					"action": "Save the appointment history."
				}
			},
			{
				"from": "caller",
				"value": {
					"tool": "save_appointment_history",
					"parameters": {
						"user_id": "JICC571413",
						"symptoms": "fatigue, chills, body aches, sore throat",
						"specialist_id": "AECJ317777",
						"appointment_time_date": "11:00-11:30, 30/11/2024"
					}
				}
			},
			{
				"from": "observation",
				"value": {
					"result": true
				}
			},
			{
				"from": "planner",
				"value": {
					"reason": "Appointment confirmed and stored.",
					"action": "Also store the current symptoms for future reference."
				}
			},
			{
				"from": "caller",
				"value": {
					"tool": "store_symptoms",
					"parameters": {
						"user_id": "JICC571413",
						"symptoms": "fatigue, chills, body aches, sore throat",
						"timestamp": "2024-09-02T10:57:00"
					}
				}
			},
			{
				"from": "observation",
				"value": {
					"result": true
				}
			},
			{
				"from": "planner",
				"value": {
					"reason": "Appointment confirmed and symptoms stored.",
					"action": "Inform the user of the successful booking."
				}
			},
			{
				"from": "caller",
				"value": {
					"tool": "notify_user",
					"parameters": {
						"user_id": "JICC571413",
						"message": "Your appointment with Dr. Diego Arroyo (General Physician) is confirmed for 11:00 AM on November 30th. Your symptoms have been recorded for future reference."
					}
				}
			},
			{
				"from": "observation",
				"value": {
					"result": true
				}
			},
			{
				"from": "planner",
				"value": {
					"reason": "The task is completed successfully.",
					"action": "<END>"
				}
			}
			]
		}
	\end{lstlisting}
	\subsection{Appointment - Reject Appointment}
	\label{app-reject-appointment-traj}
	\begin{lstlisting}
		{
			"interaction_trajectory": [
			{
				"from": "system",
				"value": {
					"user_details": {
						"user_id": "BKYA569367",
						"name": "Tomomi Gao",
						"timestamp": "2023-11-08T10:54:00"
					}
				}
			},
			{
				"from": "user",
				"value": "I've been feeling very unwell with a high fever, body aches, constant coughing, and sore throat for the past few days."
			},
			... (similar to Appointment Trajectory)
			{
				"from": "observation",
				"value": {
					"result": {
						"specialist_id": "CXAE230642",
						"name": "Dr. Diego Perez (General Practitioner)",
						"available_slot": {
							"date": "2024-11-30",
							"time": "10:00-10:30"
						}
					}
				}
			},
			{
				"from": "planner",
				"value": {
					"reason": "Dr. Diego Perez (General Practitioner) is available on the user's preferred date and time.",
					"action": "Suggest the appointment to the user and check if they want to schedule it. Dr. Diego Perez (General Practitioner) on 2024-11-30 between 10:00-10:30"
				}
			},
			{
				"from": "caller",
				"value": {
					"tool": "get_input_from_user",
					"parameters": {
						"user_id": "BKYA569367",
						"questions": "Dr. Diego Perez (General Practitioner) has an opening at 10:00 AM on November 30th. Would you like to schedule an appointment?"
					}
				}
			},
			{
				"from": "observation",
				"value": {
					"result": "No, not at this time."
				}
			},
			{
				"from": "planner",
				"value": {
					"reason": "User has declined the appointment.",
					"action": "Proceed to store the symptoms for future reference without scheduling an appointment."
				}
			},
			{
				"from": "caller",
				"value": {
					"tool": "store_symptoms",
					"parameters": {
						"user_id": "BKYA569367",
						"symptoms": "high fever, body aches, cough, sore throat",
						"timestamp": "2023-11-08T10:54:00"
					}
				}
			},
			{
				"from": "observation",
				"value": {
					"result": true
				}
			},
			{
				"from": "planner",
				"value": {
					"reason": "Symptoms stored successfully.",
					"action": "Inform the user that the symptoms have been recorded for future reference."
				}
			},
			{
				"from": "caller",
				"value": {
					"tool": "notify_user",
					"parameters": {
						"user_id": "BKYA569367",
						"message": "Your symptoms have been recorded for future reference. Please don't hesitate to reach out if you decide to see a doctor."
					}
				}
			},
			{
				"from": "observation",
				"value": {
					"result": true
				}
			},
			{
				"from": "planner",
				"value": {
					"reason": "The process is completed with symptoms stored and user notified.",
					"action": "<END>"
				}
			}
			]
		}
	\end{lstlisting}
	\subsection{Appointment - Follow-up for Unclear Symptoms}
	\label{app-followup-appointment-traj}
	\begin{lstlisting}
		{
			"interaction_trajectory": [
			{
				"from": "system",
				"value": {
					"user_details": {
						"user_id": "BLTA888285",
						"name": "Juan Martinez",
						"timestamp": "2024-08-05T04:20:00"
					}
				}
			},
			{
				"from": "user",
				"value": "I've been having some trouble with my movement lately."
			},
			{
				"from": "planner",
				"value": {
					"reason": "User reports movement issues, but I need to clarify if it's weakness, stiffness, or something else.",
					"action": "Ask user if they are experiencing weakness, stiffness, or difficulties in coordination."
				}
			},
			{
				"from": "caller",
				"value": {
					"tool": "get_input_from_user",
					"parameters": {
						"user_id": "BLTA888285",
						"questions": "Could you describe if you're feeling weakness, stiffness, or issues with coordination?"
					}
				}
			},
			{
				"from": "observation",
				"value": {
					"result": {
						"user": "I feel a bit weak and sometimes my hand shakes."
					}
				}
			},
			{
				"from": "planner",
				"value": {
					"reason": "User mentions weakness and hand shaking, which could indicate a need for rehabilitation.",
					"action": "Ask if they are experiencing any pain or if the weakness is constant."
				}
			},
			{
				"from": "caller",
				"value": {
					"tool": "get_input_from_user",
					"parameters": {
						"user_id": "BLTA888285",
						"questions": "Are you experiencing any pain, or is the weakness constant throughout the day?"
					}
				}
			},
			{
				"from": "observation",
				"value": {
					"result": {
						"user": "There's no pain, but the weakness seems to come and go."
					}
				}
			},
			.... (Rest similar to Appointment Trajectory)
			]
		}
	\end{lstlisting}
	\subsection{Soft SOS}
	\label{app-soft-sos-traj}
	\begin{lstlisting}
		{
			"interaction_trajectory": [
			{
				"from": "system",
				"value": {
					"user_details": {
						"user_id": "PKFG301655",
						"name": "Aaliyah Sousa",
						"timestamp": "2025-02-01T08:11:00"
					}
				}
			},
			{
				"from": "user",
				"value": "Soft SOS triggered. Abnormal Vitals: {'oxygen': 85, 'heart_rate': 41, 'sleep': {'deep': 75, 'light': 238, 'rem': 94, 'awake': 44}}"
			},
			{
				"from": "planner",
				"value": {
					"reason": "User has triggered a soft SOS, so the system needs to notify the user about abnormal vitals",
					"action": "Call notify_user to inform the user that the SOS process is being triggered."
				}
			},
			{
				"from": "caller",
				"value": {
					"tool": "notify_user",
					"parameters": {
						"user_id": "PKFG301655",
						"symptoms": "Soft SOS triggered. Abnormal vitals detected.\nIf you are feeling unwell, contact emergency services or book an appointment.\n\nYour Vitals-\nHeart Rate: 41 bps\nOxygen: 85%\nDeep Sleep: 75 minutes\nLight Sleep: 238 minutes\nREM: 94 minutes\nAwake: 44 minutes"
					}
				}
			},
			{
				"from": "planner",
				"value": {
					"reason": "The user has been informed about the abnormal vitals. The Soft SOS process is completed successfully.",
					"action": "<END>"
				}
			}
			]
		}
	\end{lstlisting}
	\subsection{Hard SOS - Start}
	\label{app-hard-sos-start-traj}
	\begin{lstlisting}
		{
			"interaction_trajectory": [
			{
				"from": "system",
				"value": {
					"user_details": {
						"user_id": "HNNT232992",
						"name": "Jace Cardoso",
						"timestamp": "2024-03-13T06:59:00"
					}
				}
			},
			{
				"from": "user",
				"value": "Hard SOS triggered"
			},
			{
				"from": "planner",
				"value": {
					"reason": "User has triggered a hard SOS, so the system needs to notify the user that the SOS process is starting.",
					"action": "Call notify_user to inform the user that the SOS process is being triggered."
				}
			},
			{
				"from": "caller",
				"value": {
					"tool": "notify_user",
					"parameters": {
						"user_id": "HNNT232992",
						"symptoms": "Hard SOS triggered. We are initiating emergency response procedures."
					}
				}
			},
			{
				"from": "observation",
				"value": {
					"result": true
				}
			},
			{
				"from": "planner",
				"value": {
					"reason": "The user has been notified about the SOS initiation. The system now needs to retrieve the user's location to proceed with the emergency response.",
					"action": "Call get_location to fetch the user's current coordinates."
				}
			},
			{
				"from": "caller",
				"value": {
					"tool": "get_location",
					"parameters": {}
				}
			},
			{
				"from": "observation",
				"value": {
					"result": {
						"latitude": 23.5326,
						"longitude": 139.7524
					}
				}
			},
			{
				"from": "planner",
				"value": {
					"reason": "The user's current location has been retrieved. Next, the system needs to find the nearest ambulance based on this location.",
					"action": "Call search_ambulance with the location to find the nearest available ambulance."
				}
			},
			{
				"from": "caller",
				"value": {
					"tool": "search_ambulance",
					"parameters": {
						"location": {
							"latitude": 23.5326,
							"longitude": 139.7524
						}
					}
				}
			},
			{
				"from": "observation",
				"value": {
					"result": {
						"ambulance_id": "AMBpF0E",
						"phone_no": "+146910850030"
					}
				}
			},
			{
				"from": "planner",
				"value": {
					"reason": "The nearest ambulance has been identified. The system needs to notify the ambulance about the user's location and details.",
					"action": "Call send_message to notify the ambulance with the user's location and details."
				}
			},
			{
				"from": "caller",
				"value": {
					"tool": "send_message",
					"parameters": {
						"phone_no": "+146910850030",
						"text": "Ambulance needed at location {latitude: 23.5326, longitude: 139.7524} by user HNNT232992 - Jace Cardoso"
					}
				}
			},
			{
				"from": "observation",
				"value": {
					"result": true
				}
			},
			{
				"from": "planner",
				"value": {
					"reason": "The ambulance has been notified. Next, the system needs to alert the user's emergency contacts about the SOS.",
					"action": "Call send_message to notify emergency contacts about the SOS and user's location."
				}
			},
			{
				"from": "caller",
				"value": {
					"tool": "send_message",
					"parameters": {
						"text": "SOS triggered by Jace Cardoso at location {latitude: 23.5326, longitude: 139.7524}",
						"to_emergency_contacts": true
					}
				}
			},
			{
				"from": "observation",
				"value": {
					"result": true
				}
			},
			{
				"from": "planner",
				"value": {
					"reason": "The emergency contacts have been notified. The system now needs to inform the user about the actions taken, including ambulance and emergency contact notifications.",
					"action": "Call notify_user to inform the user about the completion of the SOS process."
				}
			},
			{
				"from": "caller",
				"value": {
					"tool": "notify_user",
					"parameters": {
						"user_id": "HNNT232992",
						"symptoms": "Ambulance (AMBpF0E) with contact HNNT232992 has been informed and is on its way. Your emergency contacts have also been notified."
					}
				}
			},
			{
				"from": "observation",
				"value": {
					"result": true
				}
			},
			{
				"from": "planner",
				"value": {
					"reason": "The user has been informed about the actions taken. The SOS process is completed successfully.",
					"action": "<END>"
				}
			}
			]
		}
	\end{lstlisting}
	\subsection{Hard SOS - End}
	\label{app-hard-sos-end-traj}
	\begin{lstlisting}
		{
			"interaction_trajectory": [
			{
				"from": "system",
				"value": {
					"user_details": {
						"user_id": "EORZ618635",
						"name": "Leah Lima",
						"timestamp": "2023-11-29T01:52:00"
					}
				}
			},
			{
				"from": "user",
				"value": "End SOS triggered"
			},
			{
				"from": "planner",
				"value": {
					"reason": "User has triggered an End SOS. The system needs to notify the user that the End SOS process is being initiated.",
					"action": "Call notify_user to inform the user about the initiation of the End SOS process."
				}
			},
			{
				"from": "caller",
				"value": {
					"tool": "notify_user",
					"parameters": {
						"user_id": "EORZ618635",
						"symptoms": "End SOS triggered. We are notifying all relevant parties."
					}
				}
			},
			{
				"from": "observation",
				"value": {
					"result": true
				}
			},
			{
				"from": "planner",
				"value": {
					"reason": "The user has been informed about the initiation of the End SOS process. The system now needs to get the assigned ambulance details.",
					"action": "Call get_assigned_ambulance to retrieve the ambulance details."
				}
			},
			{
				"from": "caller",
				"value": {
					"tool": "get_assigned_ambulance",
					"parameters": {
						"user_id": "EORZ618635"
					}
				}
			},
			{
				"from": "observation",
				"value": {
					"result": {
						"ambulance_id": "AMBUaTg",
						"phone_no": "+235135781046"
					}
				}
			},
			{
				"from": "planner",
				"value": {
					"reason": "The assigned ambulance details have been retrieved. The system now needs to notify the ambulance about the SOS being ended.",
					"action": "Call send_message to inform the ambulance about the End SOS."
				}
			},
			{
				"from": "caller",
				"value": {
					"tool": "send_message",
					"parameters": {
						"phone_no": "+235135781046",
						"text": "The SOS triggered by user EORZ618635 - Leah Lima has been resolved. Ambulance services are no longer required."
					}
				}
			},
			{
				"from": "observation",
				"value": {
					"result": true
				}
			},
			{
				"from": "planner",
				"value": {
					"reason": "The ambulance has been notified about the End SOS. Next, the system needs to notify the user's emergency contacts about the resolution.",
					"action": "Call send_message to inform emergency contacts about the SOS resolution."
				}
			},
			{
				"from": "caller",
				"value": {
					"tool": "send_message",
					"parameters": {
						"text": "The SOS triggered by Leah Lima has been resolved. No further assistance is required.",
						"to_emergency_contacts": true
					}
				}
			},
			{
				"from": "observation",
				"value": {
					"result": true
				}
			},
			{
				"from": "planner",
				"value": {
					"reason": "The emergency contacts have been notified about the SOS resolution. The system now needs to inform the user that the End SOS process is completed.",
					"action": "Call notify_user to inform the user about the successful completion of the End SOS process."
				}
			},
			{
				"from": "caller",
				"value": {
					"tool": "notify_user",
					"parameters": {
						"user_id": "EORZ618635",
						"symptoms": "The SOS process has been successfully ended. Ambulance and emergency contacts have been informed."
					}
				}
			},
			{
				"from": "observation",
				"value": {
					"result": true
				}
			},
			{
				"from": "planner",
				"value": {
					"reason": "The user has been informed about the successful completion of the End SOS process. The task is completed.",
					"action": "<END>"
				}
			}
			]
		}
	\end{lstlisting}
	\section{List of Tools}
	\label{app:tools}
	Following is a list of tools for which planner and caller is trained:
	\begin{lstlisting}
		[
		{
			"Name": "get_location",
			"Description": "Fetches the GPS coordinates of the user's current location.",
			"Parameters": [],
			"Required Parameters": [],
			"Returns": {
				"type": "dictionary",
				"description": "Returns the user's current location as {latitude: float, longitude: float}."
			}
		},
		{
			"Name": "search_ambulance",
			"Description": "Finds the nearest available ambulance based on the given location.",
			"Parameters": [
			{
				"param_name": "location",
				"type": "dictionary",
				"default": "None",
				"description": "The user's current location as {latitude: float, longitude: float}."
			}
			],
			"Required Parameters": [
			"location"
			],
			"Returns": {
				"type": "dictionary",
				"description": "Returns the ambulance details {ambulance_id: string, phone_no: string}."
			}
		},
		{
			"Name": "send_message",
			"Description": "Sends a text message to a specified phone number or emergency contacts.",
			"Parameters": [
			{
				"param_name": "phone_no",
				"type": "string",
				"default": "None",
				"description": "The phone number to send the message to."
			},
			{
				"param_name": "text",
				"type": "string",
				"default": "None",
				"description": "The message text to send."
			},
			{
				"param_name": "to_emergency_contacts",
				"type": "boolean",
				"default": "False",
				"description": "Whether to send the message to all emergency contacts."
			}
			],
			"Required Parameters": [
			"text"
			],
			"Returns": {
				"type": "boolean",
				"description": "Returns True if the message was sent successfully."
			}
		},
		{
			"Name": "get_available_specialists",
			"Description": "Fetches a list of specialists and their availability.",
			"Parameters": [
			{
				"param_name": "symptoms",
				"type": "string",
				"default": "None",
				"description": "List of symptoms derived from user input."
			},
			{
				"param_name": "specialization",
				"type": "string",
				"default": "None",
				"description": "Specialization name for the appointment."
			},
			{
				"param_name": "user_schedule",
				"type": "object",
				"default": "None",
				"description": "User's preferred schedule for appointments in key-value pairs (e.g., {'date': 'YYYY-MM-DD', 'time_range': 'HH:MM-HH:MM'})."
			}
			],
			"Required Parameters": [
			"symptoms",
			"user_schedule"
			],
			"Returns": {
				"type": "dictionary",
				"description": "Returns single best schedule {specialist_id: string, name: string, available_slot: string, date: string}."
			}
		},
		{
			"Name": "confirm_appointment",
			"Description": "Confirms an appointment slot and stores it in the hospital's database.",
			"Parameters": [
			{
				"param_name": "user_id",
				"type": "string",
				"default": "None",
				"description": "Unique identifier for the user."
			},
			{
				"param_name": "specialist_id",
				"type": "string",
				"default": "None",
				"description": "Unique identifier for the chosen specialist."
			},
			{
				"param_name": "appointment_time_date",
				"type": "string",
				"default": "None",
				"description": "The selected time slot for the appointment as time in HH:MM-HH:MM format and date in DD/MM/YY format."
			}
			],
			"Required Parameters": [
			"user_id",
			"specialist_id",
			"appointment_time_date"
			],
			"Returns": {
				"type": "boolean",
				"description": "Always returns True."
			}
		},
		{
			"Name": "save_appointment_history",
			"Description": "Saves appointment information in the user's database for later reference and recurring use cases.",
			"Parameters": [
			{
				"param_name": "user_id",
				"type": "string",
				"default": "None",
				"description": "Unique identifier for the user."
			},
			{
				"param_name": "symptoms",
				"type": "string",
				"default": "None",
				"description": "The symptoms described by the user."
			},
			{
				"param_name": "specialist_id",
				"type": "string",
				"default": "None",
				"description": "Unique identifier for the chosen specialist."
			},
			{
				"param_name": "appointment_time_date",
				"type": "string",
				"default": "None",
				"description": "The selected time slot for the appointment as time in HH:MM-HH:MM format and date in DD/MM/YY format."
			}
			],
			"Required Parameters": [
			"user_id",
			"symptoms",
			"specialist_id",
			"appointment_time_date"
			],
			"Returns": {
				"type": "boolean",
				"description": "Always returns True."
			}
		},
		{
			"Name": "get_available_specialists",
			"Description": "Fetches a list of specialists and their availability.",
			"Parameters": [
			{
				"param_name": "symptoms",
				"type": "string",
				"default": "None",
				"description": "List of symptoms derived from user input."
			},
			{
				"param_name": "specialization",
				"type": "string",
				"default": "None",
				"description": "specialization name for the appointment"
			},
			{
				"param_name": "user_schedule",
				"type": "object",
				"default": "None",
				"description": "User's preferred schedule for appointments in key-value pairs (e.g., {'date': 'YYYY-MM-DD', 'time_range': 'HH:MM-HH:MM'})."
			}
			],
			"Required Parameters": [
			"symptoms",
			"specialization"
			],
			"Returns": {
				"type": "dictionary",
				"description": " returns single best schedule {specialist_id,name, available_slot including time in HH:MM-HH:MM format and date in DD/MM/YY format}."
			}
		},
		{
			"Name": "confirm_appointment",
			"Description": "Confirms an appointment slot and stores it in the hospital's database.",
			"Parameters": [
			{
				"param_name": "user_id",
				"type": "string",
				"default": "None",
				"description": "Unique identifier for the user."
			},
			{
				"param_name": "specialist_id",
				"type": "string",
				"default": "None",
				"description": "Unique identifier for the chosen specialist."
			},
			{
				"param_name": "appointment_time_date",
				"type": "string",
				"default": "None",
				"description": "The selected time slot for the appointment as time in HH:MM-HH:MM format and date in DD/MM/YY format."
			}
			],
			"Required Parameters": [
			"user_id",
			"specialist_id",
			"appointment_time"
			],
			"Returns": {
				"type": "boolean",
				"description": "Always returns True."
			}
		},
		{
			"Name": "save_appointment_history",
			"Description": "Saves appointment information in the user's database for later reference and recurring use cases.",
			"Parameters": [
			{
				"param_name": "user_id",
				"type": "string",
				"default": "None",
				"description": "Unique identifier for the user."
			},
			{
				"param_name": "symptoms",
				"type": "string",
				"default": "None",
				"description": "The symptoms described by the user."
			},
			{
				"param_name": "specialist_id",
				"type": "string",
				"default": "None",
				"description": "Unique identifier for the chosen specialist."
			},
			{
				"param_name": "appointment_time_date",
				"type": "string",
				"default": "None",
				"description": "The selected time slot for the appointment as time in HH:MM-HH:MM format and date in DD/MM/YY format."
			}
			],
			"Required Parameters": [
			"user_id",
			"symptoms",
			"specialist_id",
			"appointment_time_date"
			],
			"Returns": {
				"type": "boolean",
				"description": "Always true"
			}
		},
		{
			"Name": "get_appointment_history",
			"Description": "Retrieves the user's appointment history for analysis and reminders.",
			"Parameters": [
			{
				"param_name": "user_id",
				"type": "string",
				"default": "None",
				"description": "Unique identifier for the user."
			}
			],
			"Required Parameters": [
			"user_id"
			],
			"Returns": {
				"type": "array",
				"description": "Array containing past appointment records."
			}
		},
		{
			"Name": "retrieve_past_complaints",
			"Description": "Fetches the user's past complaints matching the given symptoms for analysis and reference.",
			"Parameters": [
			{
				"param_name": "user_id",
				"type": "string",
				"default": "None",
				"description": "Unique identifier for the user."
			},
			{
				"param_name": "symptoms",
				"type": "string",
				"default": "None",
				"description": "List of symptoms to search for in past complaints."
			},
			{
				"param_name": "date_range",
				"type": "object",
				"default": "None",
				"description": "Optional date range filter in the format {'start_date': 'YYYY-MM-DD', 'end_date': 'YYYY-MM-DD'}."
			}
			],
			"Required Parameters": [
			"user_id",
			"symptoms"
			],
			"Returns": {
				"type": "array",
				"description": "Array of past complaints related to the specified symptoms."
			}
		},
		{
			"Name": "follow_up_with_user",
			"Description": "Initiates a follow-up interaction with the user based on their past complaints and current symptoms.",
			"Parameters": [
			{
				"param_name": "user_id",
				"type": "string",
				"default": "None",
				"description": "Unique identifier for the user."
			},
			{
				"param_name": "past_complaints",
				"type": "array",
				"default": "[]",
				"description": "List of past complaints to reference during the follow-up."
			},
			{
				"param_name": "current_symptoms",
				"type": "string",
				"default": "None",
				"description": "Current symptoms reported by the user."
			},
			{
				"param_name": "preferred_contact_method",
				"type": "string",
				"default": "None",
				"description": "User's preferred method for follow-up (e.g., 'call', 'email', 'chat')."
			}
			],
			"Required Parameters": [
			"user_id",
			"current_symptoms"
			],
			"Returns": {
				"type": "object",
				"description": "Details of the follow-up initiated, including method and next steps."
			}
		},
		{
			"Name": "notify_user",
			"Description": "Sends a notification to the user ",
			"Parameters": [
			{
				"param_name": "user_id",
				"type": "string",
				"default": "None",
				"description": "Unique identifier for the user."
			},
			{
				"param_name": "message",
				"type": "string",
				"default": "None",
				"description": "The notification message to send."
			}
			],
			"Required Parameters": [
			"user_id",
			"message"
			],
			"Returns": {
				"type": "boolean",
				"description": "Always returns status as True"
			}
		},
		{
			"Name": "get_input_from_user",
			"Description": "Collects input from the user for specified parameters.",
			"Parameters": [
			{
				"param_name": "user_id",
				"type": "string",
				"default": "None",
				"description": "Unique identifier for the user."
			},
			{
				"param_name": "questions",
				"type": "string",
				"default": "[]",
				"description": "Question to ask the user."
			}
			],
			"Required Parameters": [
			"user_id",
			"questions"
			],
			"Returns": {
				"type": "string",
				"description": "User's response to the specified questions."
			}
		},
		{
			"Name": "store_symptoms",
			"Description": "Stores the symptoms reported by the user and initiates a follow-up process to gather additional details for a more accurate analysis.",
			"Parameters": [
			{
				"param_name": "user_id",
				"type": "string",
				"default": "None",
				"description": "Unique identifier for the user."
			},
			{
				"param_name": "symptoms",
				"type": "string",
				"default": "None",
				"description": "Symptoms of user."
			},
			{
				"param_name": "timestamp",
				"type": "object",
				"default": "None",
				"description": "Save date and time of the event",
				"Required Parameters": [
				"user_id",
				"symptoms",
				"timestamp"
				],
				"Returns": {
					"type": "boolean",
					"description": "Status always true"
				}
			}
			]
		}
		]
	\end{lstlisting}
	
\end{document}